\documentclass{ifacconf}

\usepackage{amssymb}
\usepackage{bm}
\usepackage{amsmath}
\usepackage{graphicx}
\usepackage{xcolor}
\usepackage{mathtools}
\usepackage{natbib}

\begin{document}
\begin{frontmatter}

\title{Locating fast-varying line disturbances with the frequency mismatch}

\thanks[footnoteinfo]{RD was supported by the Swiss National Science Foundation under grant P400P2\_194359.}

\author{Robin Delabays,\textsuperscript{1} Laurent Pagnier,\textsuperscript{2} and Melvyn Tyloo\textsuperscript{3}} 

\address{$^1$ Center for Control, Dynamical Systems, and Computation, UCSB, \\ Santa Barbara, CA 93106, USA.\\
$^2$ Department of Mathematics, University of Arizona, \\ Tucson, AZ 85721, USA. \\
$^3$ Theoretical Division, Los Alamos National Laboratory, \\ Los Alamos, NM 87545, USA.}

\begin{abstract}
In an attempt to provide an efficient method for line disturbance identification in complex networks of diffusively coupled agents, we recently proposed to leverage the \emph{frequency mismatch}. 
The frequency mismatch filters out the intricate combination of interactions induced by the network structure and quantifies to what extent the trajectory of each agent is affected by the disturbance. In this previous work, we provided some analytical evidence of its efficiency when the perturbation is assumed to be \emph{slow}.
In the present work, we claim that the frequency mismatch performs actually well for most disturbance regimes. 
This is shown through a series of simulations and is backed up by an analytical argument. 
Therefore, we argue that the frequency mismatch is an efficient and elegant tool for line disturbance location in complex networks of diffusively coupled agents. 

\vspace{5mm}

Copyright 2022 The Authors. This work has been accepted to IFAC for publication under a Creative Commons Licence CC-BY-NC-ND.
\end{abstract}


\end{frontmatter}

\section{Introduction}
Networked system models find numerous applications in the description of natural and engineered systems~[\cite{strogatz2001exploring, pikovski2003synchronization}]. 
Following from this ubiquity, network science has grown as a natural link between fields as diverse as pure mathematics (with graph theory), neuroscience (with the interactions patterns in the brain), and engineering (e.g., in the context of power grids). 

The various components of a networked system are naturally subjected to a diversity of disturbances, ranging from the natural environmental noise, unexpected defects, to the major breakdowns due to extreme events or attacks. 
In engineered systems in particular, the identification of these disturbances, e.g., based on measurements, is key to a safe operation. 
Whereas detecting a disturbance can be done by comparing the actual trajectory of the system with its expected trajectory, locating the disturbance can be much more difficult to tackle, particularly in complex networked systems. 

Moreover, in a network, disturbances can occur either on nodes or on edges. 
While nodal disturbances usually intervene additively in the models, line disturbances are typically multiplicative, rendering them significantly harder to track analytically. 

Models of networked systems often describe the diffusion of a commodity or information among a group of agents, the couplings are therefore referred to as \emph{diffusive}. 
In normal operation, such systems typically evolve towards a steady state, e.g., consider voltage dynamics in power grids~[\cite{bergen1981structure,dorfler13synchronization}] or opinion dynamics over social networks~[\cite{castellano2009statistical}], to name just two examples. 
Therefore, disturbances typically drive the system away from its steady state. 

Assuming that the disturbance does not completely impede the operation of the system, analyzing the system's response to the perturbation can provide valuable information about the system itself, as well as on the perturbation. 
For instance, comparing the arrival time of a disturbance at different points of the network allows to perform a triangulation, pointing to the source~[\cite{semerow2016disturbance}]. 
The \emph{Discrete Wavelength Transform} has also proved to be able to locate the source of a disturbance~[\cite{upadhyaya2015power,mathew2016pmu}], but is limited to nodal disturbances. 
More closely related to our interest here, some approaches have been developed specifically for the identification of line disturbances. 
While some approaches are analytical by nature~[\cite{coletta2018performance}], most of them rely on an optimization scheme, trying to find the best set of disturbed lines that explains the measurements~[\cite{soltan2017analyzing,soltan2018power,soltan2019expose,jamei2020phasor}]. 

In~\cite{delabays2021locating}, we proposed to use the \emph{frequency mismatch} (see below for a formal definition) to identify the two ends of the disturbed line, in a network of diffusively coupled agents. 
This approach is backed up by analytical evidence that requires the disturbance to occur sufficiently slowly.
Namely, we assumed that the characteristic time of the disturbance was smaller than the system's characteristic times. 
Intuitively, the system adapts its steady state as the disturbance kicks in. 

We show here that the frequency mismatch allows to identify the disturbed line for much faster disturbances than what is assumed in~\cite{delabays2021locating}. 
For cases of practical interest, we even observe that the frequency mismatch approach performs perfectly, independently of the disturbance's characteristic time. 
We support our numerical observations with an analytic rationale.

\section{The model}
Let first recall the setup and results of \cite{delabays2021locating}. 
We consider a set of $n$ diffusively, network-coupled agents, whose states $x_i\in\mathbb{R}$ are governed by the differential equations,
\begin{align}\label{eq:dyn}
m_i\,\ddot{x}_i + d_i\,\dot{x}_i &= \omega_i - \sum_j a_{ij}f(x_i-x_j)\, , & i &= 1,...,n\, ,
\end{align}
where $d_i$ is the damping, $m_i$ the inertia, and $\omega_i$ the natural velocity of agent $i$. 
Without loss of generality, we assume $\bm{\omega}^\top\bm{1}=0$, considering the variables in a moving/rotating frame if necessary. 
The elements of the adjacency matrix of the coupling network are denoted $a_{ij}$ and are taken symmetric ($a_{ij}=a_{ji}$). 
The coupling function $f$ is assumed to be odd and differentiable with $f'(0)>0$, guaranteeing it is attractive in a neighborhood of the origin. 

Let $\bm{x}^*\in\mathbb{R}^n$ be a stable steady state of Eq.~\eqref{eq:dyn}. 
Provided the system is not too heterogeneous, i.e., the spread of the natural frequencies $\max_i\omega_i-\min_i\omega_i$) is moderate relative to the coupling strengths ($a_{ij}$), the steady state $\bm{x}^*$ is reasonably estimated by the solution of the linearized equation 
\begin{align}\label{eq:jac}
 \bm{\omega} &= L_0 \bm{x}\, , & L_{0,ij} &= -\frac{\partial F_i(\bm{0})}{\partial x_j} = -w_{ij}\, ,
\end{align}
where $F_i$ is the sum in the right-hand-side of Eq.~\eqref{eq:dyn}. 
The matrix $L_0$ can be seen as the Laplacian matrix of the interaction graph, with weights given by Eq.~\eqref{eq:jac}. 

{\bf Remark.}
 {\it For various relevant applications (e.g., power grids), the assumptions above are quite standard [\cite{machowski2008power}]. 
 In summary, we require that, close to its stable fixed point, the system is well approximated by a diffusive linear time-invariant system. 
 Indeed, if the system in Eq.~(\ref{eq:dyn}) is already diffusive linear time-invariant system, these assumptions are trivially satisfied.}

Under the assumption that the coupling functions are increasing and that the graph is connected, the matrix $L_0$ has one vanishing eigenvalue and all others are positive [\cite{fiedler1973algebraic}], 
\begin{align}
 0 &= \lambda_1 < \lambda_2 \leq \cdots \leq \lambda_n\, .
\end{align}
The nonzero eigenvalues describe the linear behavior of the system in the neighborhood of the origin. 
Namely, the time needed by the system to relax to its steady state is given by $\sim\lambda_2^{-1}[{\rm s}]$. 
Therefore, under a disturbance whose rate of change is lower than $\lambda_2[{\rm s}^{-1}]$, the whole system will progressively adapt to the disturbance, whose impact will spread throughout the whole system. 
On the other side of the spectrum, a disturbance whose rate of change is larger than $\lambda_n[{\rm s}^{-1}]$ will vary too fast for its influence to spread, and will be limited to the disturbed components of the system.

\section{Line disturbance and the frequency mismatch}
Let us assume that line $(i,j)$ is perturbed. 
The coupling $a_{ij}$ then becomes time-dependent, which we model in Eq.~\eqref{eq:dyn} as, 
\begin{align}
 a_{ij}(t) &= a^0_{ij} + \xi_{\rm l}(t)\, ,
\end{align}
with $a^0_{ij}$ the vase value and $\xi_{\rm l}$ being the actual line disturbance. 

For sake of convenience, we take a sinusoidal perturbation, 
\begin{align}
 \xi_{\rm l}(t) &= \xi_0\sin(\Omega t)\, ,
\end{align}
but our approach is more general. 
Such a sinusoidal disturbance has the advantage of clearly emphasizing the time scale of the perturbation, here $\Omega \geq 0$, which can easily be tuned in simulations.

{\bf Slow perturbations.} 
Under slow perturbation, i.e., $\Omega < \lambda_2$, the system progressively follows the disturbance. 
Therefore, we estimate the response of the system as the solution of Eq.~\eqref{eq:jac}, under a time varying Laplacian matrix. 
Namely, 
\begin{align}\label{eq:slow}
 \bm{x}(t) &\approx [L_0 + \xi(t)\bm{e}_{ij}\bm{e}_{ij}^\top]^\dagger \bm{\omega}\, ,
\end{align}
where $\dagger$ denotes the Moore-Penrose pseudoinverse, $\bm{e}_{ij}=\bm{e}_i-\bm{e}_j$, and $\bm{e}_i$ is the $i^{\rm th}$ vector of the canonical basis. 
Eq.~\eqref{eq:slow} holds because we assumed $\bm{\omega}$ to be in the null-space of $L_0$. 

In \cite{delabays2021locating}, we defined the \emph{frequency mismatch} 
\begin{align}\label{eq:psi}
 \bm{\psi}(t) &= L_0 \left[\bm{x}(t) - \bm{x}^*\right]\, ,
\end{align}
as a mean to identify the two ends of the perturbed lines, in the case where the rate of change of the perturbation is sufficiently slow. 
The matrix $L_0$ is known if we know the system under investigation, which is a prerequisite of the frequency mismatch approach. 

The perturbed line is inferred as being the link between the two nodes whose trajectories have the largest amplitudes in $\bm{\psi}(t)$. 
Formally, we have
\begin{align}\label{eq:amplitudep}
 \eta_i &= \max_{t\geq 0} \psi_i(t) - \min_{t\geq 0} \psi_i(t)\, , & i &\in \{1,...,n\}\, ,
\end{align}
the frequency mismatch amplitude at node $i$. 
We refer to this approach as the \emph{$\bm{\psi}$-based approach}. 

\begin{figure*}
 \centering
 \includegraphics[width=\textwidth]{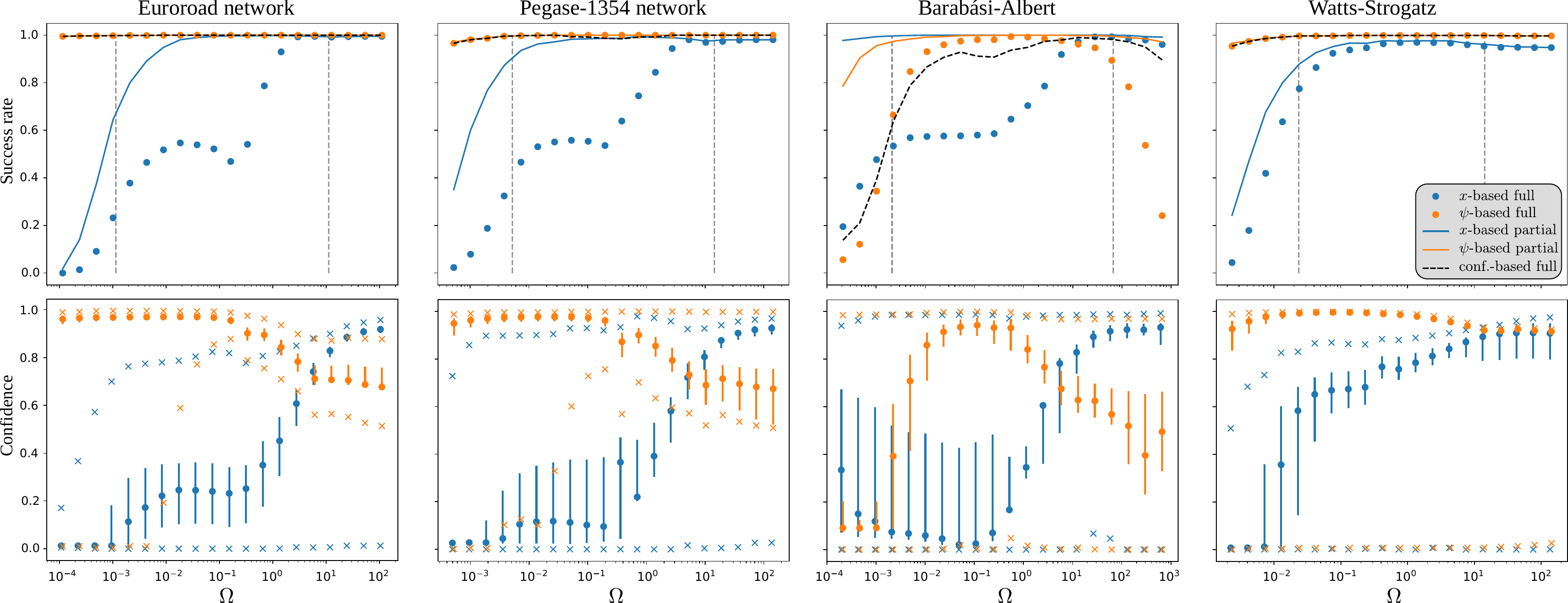}
 \caption{Top row: Success rate of the two methods of line perturbation inference ($\bm{x}$-based in blue and $\bm{\psi}$-based in orange), on four different network structures. 
 Bullets: Proportion of simulations (over $1000$) where the two ends of the line are correctly identified. 
 Plain line: Proportion of the simulations where at least one end of the line is identified. 
 The black dashed curve is the success rate when inference is based on both method, taking the inference of the method with highest confidence. 
 The black vertical lines are the eigenvalues $\lambda_2$ and $\lambda_n$ respectively. 
 For $\Omega<\lambda_2$ (resp. $\Omega>\lambda_n$) the perturbation can be considered as "slow" (resp. "fast") with respect of the system's time scales. 
 Perturbations with time scales in the interval between the two lines are "intermediate". 
 Bottom row: Statistics of the confidence estimate over the $1000$ line perturbations [Eq.~(\ref{eq:confidence})]. 
 The bullet is the median, the bar covers the 2nd and 3rd quartiles, and the crosses are the extreme values. 
 From left to right, the networks considered are: the Euroroad network~[\cite{subelj2011robust,peixoto2020the}]; the IEEE Pegase-1354 test case~[\cite{fliscounakis2013contingency}]; a Barab\'asi-Albert random network~[\cite{barabasi1999emergence}]; and a Watts-Strogatz random network~[\cite{watts1998collective}].
 }
 \label{fig:eff_om}
\end{figure*}

{\bf Fast perturbation.}
When the perturbation is fast, i.e., $\Omega > \lambda_n$, it can be described by looking at the effect of a sudden delta disturbance on a line, i.e., $L_0 \rightarrow L_0+\xi_0\delta(t)\bm{e}_{ij}\bm{e}_{ij}^\top$, where $\delta(t)$ is the standard notation for the Dirac-delta. 
Then calculating the short time response of Eq.~(\ref{eq:dyn}) following the disturbance in the inertialess case yields,
\begin{align}
 \bm{x}(t) &\cong \bm{x}^* + \int_0^t \left[L_0 + \xi_0\delta(t')\bm{e}_{ij}\bm{e}_{ij}^\top\right]\bm{x}(t')dt' \notag \\
 &= \bm{x}^* + \xi_0 \left(x_i^*-x_j^*\right)\bm{e}_{ij} + \mathcal{O}(t)\, . 
\end{align}
The latter expression shows that, at short time, only the two ends of the disturbed line react, and they do with equal amplitudes. 

A fast varying signal can naturally be approximated by a series of delta disturbances, 
\begin{align}
 \xi(t) &= \sum_k \xi_k \delta(t-k\tau)\, , & \tau	 &\ll \lambda_n^{-1}\, .
\end{align}
It is then reasonable to write the time series of the nodes as
\begin{align}\label{eq:x}
 \bm{x}(t) &= \bm{x}^* + \alpha(t)(\bm{e}_i - \bm{e}_j)\, ,
\end{align}
with $\bm{x}^*$ being the base case and $\alpha(t)$ gathering all variation of the time series
\begin{align}
 \alpha(t) &= \sum_{k\colon k\tau < t} \xi_k\, .
\end{align}
Therefore, a fast perturbation that has zero time-average does not propagate throughout the network. 

In this case, the perturbed line is inferred similarly as in the slow disturbance case, but using the trajectory $\bm{x}(t)$ instead of the frequency mismatch $\bm{\psi}(t)$.
Formally, the amplitude of the signal at node $i$ is 
\begin{align}\label{eq:amplitude}
 \eta'_i &= \max_{t\geq 0} x_i(t) - \min_{t\geq 0} x_i(t)\, , & i &\in \{1,...,n\}\, .
\end{align}
This approach is the \emph{$\bm{x}$-based approach}.

\section{Intermediate perturbations}
According to the discussion in \cite{delabays2021locating}, one would expect that the frequency mismatch $\bm{\psi}(t)$ cannot be used if the disturbance is fast with respect to the systems intrinsic time scales. 
In such a case, one would naturally rely on the actual agents' trajectories $\bm{x}(t)$, which, according to Eq.~\eqref{eq:x}, should clearly identify the two ends of the perturbed line. 
However, if the disturbance has an intermediate time scale, i.e., $\lambda_2\leq\Omega\leq\lambda_n$, then one would hope that at least one of the two above indicators performs reasonably well in the regime, but without proper guarantees. 

{\bf Numerical evidence.}
In an attempt to investigate this question, we simulated line disturbances in four different networks of Kuramoto oscillators, with 1$^{\rm st}$ order dynamics ($m_i=0$, $d_i=1$, $f=\sin$, $i=1,...,n$). 
In order to cover a diversity of network types, we considered both realistic networks and random networks:
\begin{description}
 \item[Euroroad network:] Representation of the European road network~[\cite{subelj2011robust,peixoto2020the}], composed of $n=1039$ nodes and $m=1305$ lines;
 \item[Pegase 1354:] Portion of the European power grid~[\cite{fliscounakis2013contingency}], composed of $n=1354$ nodes and $m=1710$ lines;
 \item[Barab\'asi-Albert:] A realization of a Barab\'asi-Albert network~[\cite{barabasi1999emergence}], with $n=1200$ nodes, each connected to one other node, i.e., $m=1199$;
 \item[Watts-Strogatz:] A realization of a Watts-Strogatz network~[\cite{watts1998collective}], with $n=1200$ nodes and $m=6000$, and each edge is rewired with probability $p=0.01$ (small-world regime).`
\end{description}
For each network, we simulated the sinusoidal perturbation of $1000$ distinct, randomly chosen lines, over $1000$ time steps.
Simulations were initialized from the steady state associated to a randomly chosen vector of natural frequencies $\bm{\omega}$. 
In each realization, the time step size was chosen to be either $h=0.01[{\rm s}]$ or one tenth of a disturbance period, $h=(10\Omega)^{-1}$, whichever is smaller. 

For each system, we computed the rate of success of both approaches in the task of identifying the disturbed line, which we show in Fig.~\ref{fig:eff_om} (top row). 
As expected, for fast perturbations, the disturbed line is clearly identified by the direct trajectory of the system (blue dots), and frequency mismatch performs (almost) perfectly well for slow perturbations (orange dots).

Interestingly, we systematically observe that the frequency mismatch performs very well for intermediate perturbation frequencies, and even for fast perturbations in most cases.
Even though this is quite surprising at first sight, we provide an analytical intuition to this observation. 

For the Barab\'asi-Albert network, the poor performance at low frequency is due to the limited length of the time series. 
Indeed, we see that in this case, the low frequency regime has very small values of $\Omega$, meaning that long time series are needed to observe a significant impact of the perturbation. 
Actually, we see such a decrease in success rate in all networks, for disturbance frequency below $0.01[{\rm s}^{-1}]$. 
For such low frequency, the $1000$ time steps represent a tiny fraction of a disturbance period, e.g., for frequency $\Omega=0.01$ and time step $h=0.01$, the simulation represents less than $2\%$ of a period. 
It is therefore remarkable that the $\bm{\psi}$-based disturbance location has a success rate larger than $95\%$. 
We observed (but did not show here) that increasing the length of the simulation time increased the success rate of the identification for all networks.  

\begin{figure}
 \centering
 \includegraphics[width=.95\columnwidth]{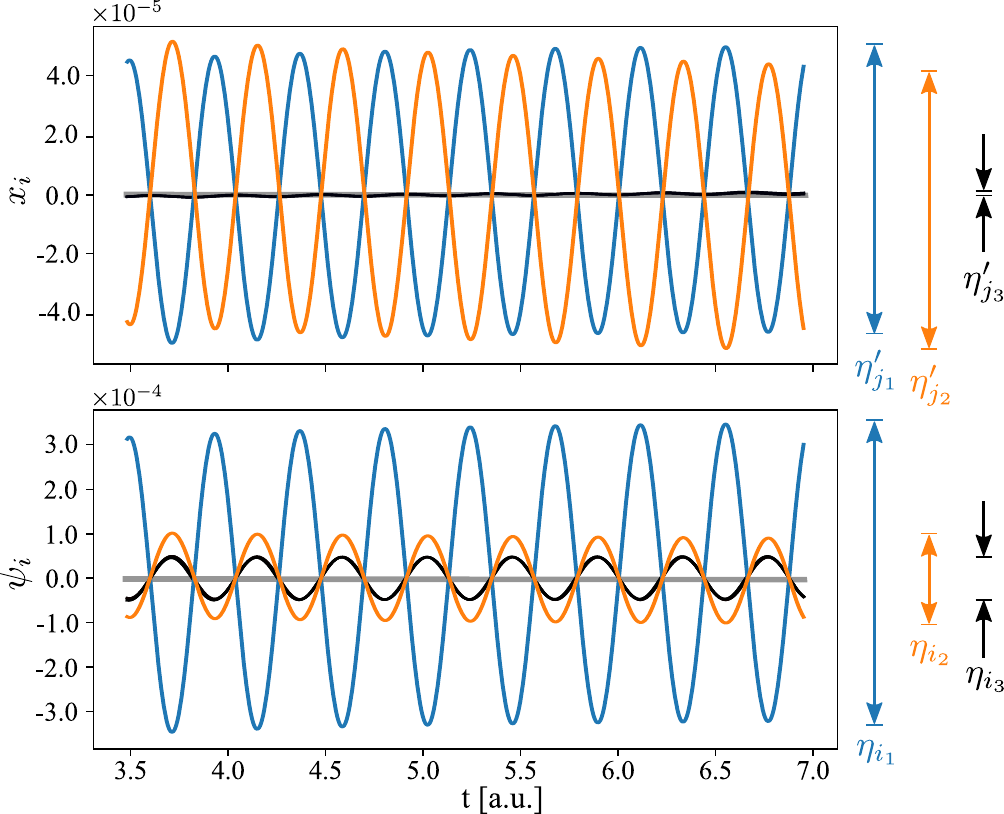}
 \caption{Trajectories of $\bm{x}$ and $\bm{\psi}$ for a group of Kuramoto oscillators~[\cite{kuramoto1984cooperative}], interacting according to the network of the IEEE Pegase-1354 test case~[\cite{fliscounakis2013contingency}]. 
 The line between buses 88 and 93 is perturbed with a frequency $\Omega = 10\lambda_n [{\rm s}^{-1}]$ and unit amplitude. 
 Bus 93 (orange curve) has a unique neighbor (bus 88) and bus 88 (blue curve) has five other neighbors. 
 This is why there are five other components of $\bm{\psi}$ that are triggered in the bottom panel (black curves,  overlapping). 
 The agents trajectories (top panel) identify very clearly the two ends of the line, but the frequency mismatch (bottom panel) does a very reasonable job as well. 
 The amplitude of the trajectory of $\psi_{93}$ is roughly twice as large as the other neighbors of node 88, as is expected from Eq.~(\ref{eq:psi_comp}). 
 The main difference between the outcomes of the two methods lies in the confidence estimates, based on the amplitudes shown on the right. 
 We have $c_{\bm x}\approx 0.98$ and $c_{\bm\psi}\approx 0.51$. }
 \label{fig:xNpsi}
\end{figure}

{\bf Analytical evidence.}
Let us compute the frequency mismatch $\bm{\psi}(t)$, under the fast perturbation case, i.e., $\bm{x}(t)$ is given by Eq.~\eqref{eq:x}, 
\begin{align}\label{eq:psi_comp}
 \psi_k(t) &= \alpha(t) \left(L\bm{e}_i - L\bm{e}_j\right)_k = \alpha(t)\left(L_{ki} - L_{kj}\right) \notag \\
 &= \left\{
 \begin{array}{ll}
  \alpha(t) ({\rm deg}_i + w_{ij})\, , & \text{if } k=i\, , \\
  \alpha(t) (-w_{ij} - {\rm deg}_j)\, , & \text{if } k=j\, , \\
  \alpha(t)(w_{kj} - w_{ki})\, , & \text{otherwise.}
 \end{array}
 \right.
\end{align}
Notice that $\psi_k(t)$ vanishes if $k$ is neither a neighbor of $i$ nor $j$, because in this case $w_{ki}=w_{kj}=0$. 
It clearly appears that, most of the time, the amplitude of $\bm{\psi}(t)$ at the two ends of the perturbed line is of the same order of magnitude as the degree of the node. 
Therefore, provided the weights are not too heterogeneous, the magnitudes at the two ends of the perturbed line will dominate the magnitude of all other components of $\bm{\psi}(t)$ which are at most of the order of the weight of a single edge. 
Even when one end of the disturbed line (say node $i$) is a leaf of the graph, one sees that its component's amplitude $\psi_i(t)$ is twice as large as those of the other neighbors of of $j$. 
All elements of this discussion are illustrated in Fig.~\ref{fig:xNpsi}. 

Confusion in the $\bm{\psi}$-based inference may arise in at least two cases. 
First, when the edge weights are too heterogeneous, the above argument breaks down and the inference might fail. 

Second, when the difference in degrees between nodes $i$ and $j$ is very large, as can happen in scale-free networks. 
We observe such a confusion in Fig.~\ref{fig:eff_om}, for the Barab\'asi-Albert network with large disturbance frequencies. 
By construction, the Barab\'asi-Albert network is scale free, meaning that there are a few \emph{hubs}, i.e., nodes with large degree, and a lot of \emph{peripheral} nodes, that have low degree (often equal to one). 
Therefore, there is a significant number of lines that match the example shown in Fig.~\ref{fig:xNpsi}, but with the high-degree node having a much larger degree. 
In such a case, the second and third largest amplitudes are almost indistinguishable, explaining the recurring mistake in line inference seen in Fig.~\ref{fig:eff_om} for the Barab\'asi-Albert network. 

{\bf Inference confidence.}
The two panels of Fig.~\ref{fig:xNpsi} clearly emphasize that, even though both methods unambiguously identify the two end-nodes of the perturbed line, relying on the trajectories $\bm{x}(t)$ (top panel) seems more trustworthy. 
To make this more precise, we propose to quantify the confidence in the inference outcome of each method. 
To do so, let us define the two following orderings of the node indices, $(i_1,i_2,...,i_n)$ and $(j_1,j_2,...,j_n)$ such that
\begin{align}\label{eq:ordering}
 \eta_{i_1} &\geq \eta_{i_2} \geq \cdots \geq \eta_{i_n}\, , &
 \eta_{j_1}' &\geq \eta_{j_2}' \geq \cdots \geq \eta_{j_n}'\, ,
\end{align}
with amplitudes defined in Eqs.~\eqref{eq:amplitudep} and \eqref{eq:amplitude}. 
The perturbed line inferred by the $\bm{\psi}$-based (resp. $\bm{x}$-based) approach is then $(i_1,i_2)$ [resp. $(j_1,j_2)$]. 

We define the \emph{confidence estimate} of the inferred location as 
\begin{align}\label{eq:confidence}
 c_{\bm \psi} &= 1 - \eta_{i_3}/\eta_{i_2}\, , & c_{\bm x} &= 1 - \eta_{j_3}'/\eta_{j_2}'\, .
\end{align}
We note that $c_{\bm \psi}, c_{\bm x} \in [0,1]$.
Indeed, large values of the confidence mean that there is a large relative gap between $\eta_{i_2}$ and $\eta_{i_3}$ (resp. $\eta_{j_2}'$ and $\eta_{j_3}'$), and therefore the estimate is very clear. 
Whereas, if the confidence is low, this means that the second and third largest amplitudes are very similar and a mistake is more likely in the inference. 
Statistics of the confidence for our four test networks are given in the bottom panel of Fig.~\ref{fig:eff_om}, emphasizing the switching from high $c_{\bm \psi}$ to high $c_{\bm x}$ (except for Barab\'si-Albert, which has already been discussed). 

With a confidence estimate at hand, it is then tempting to base the decision of which method to use upon its self-estimated confidence. 
In the top row of Fig.~\ref{fig:eff_om}, the dashed black line shows the success rate of the inference where the method used is the one that has the largest confidence. 
As can be expected, the confidence-based estimate cannot perform better than the best of the two methods. 
Indeed, it happens that one of the methods makes a wrong inference with high confidence, hence decreasing the performance of the confidence-based estimate. 
Nevertheless, in the only regime where the frequency mismatch does not perform well (Barab\'asi-Albert, large frequency), the confidence-based approach copes with this inefficiency. 
There is then a trade-off to find between a very efficient method that fails in some cases and a slightly less successful method on average, but which minimizes the worst error. 

\begin{figure}
\includegraphics[width=\columnwidth]{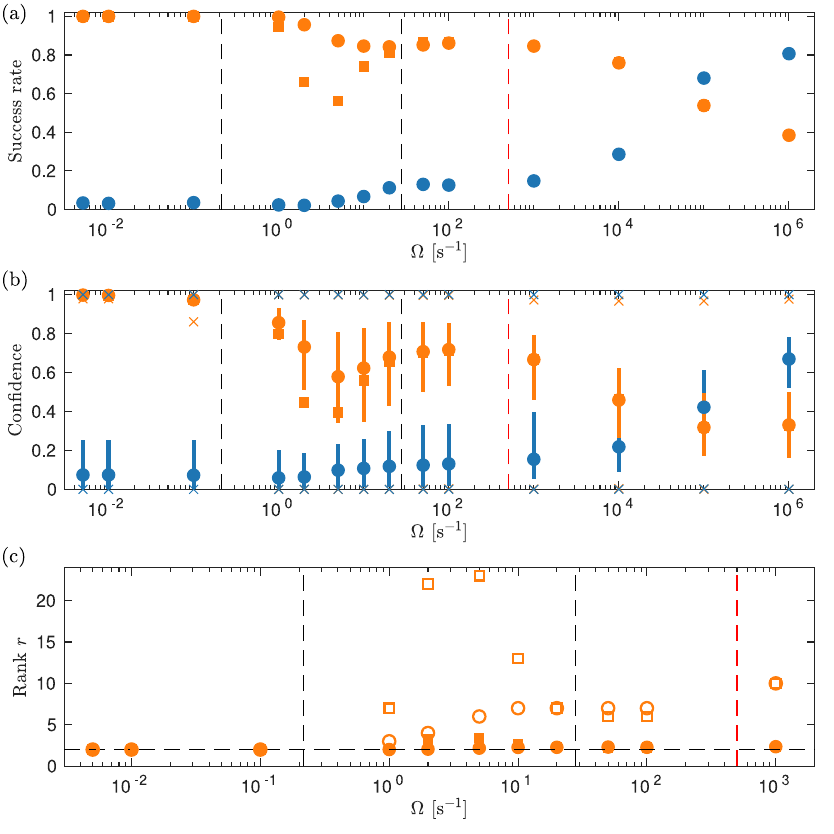}
\caption{(a) Success rate of the line perturbation inference methods: $\bm{x}$-based (blue) and $\bm{\psi}$-based (orange). 
For the $\bm{\psi}$-based  method, we distinguish early and late detection which are denoted by bullets and squares respectively. 
(b) Confidence estimates defined in Eq.~\eqref{eq:confidence}. 
(c) Rank detection $r$. 
Filled markers denotes the average rank and hollow ones the maximal rank attained over all disturbed lines. 
In all panels, the dashed red line represents the 10$^{\rm th}$ harmonic of the grid frequency (i.e., 500Hz) and the dashed black lines indicate the range over which the eigenvalues of the system are spread. 
Disturbances were applied to 504 transformers in the continental European grid. }
\label{fig:pantagruel}
\end{figure}

{\bf Realistic power grids.}
We previously tested our method on synthetic networks with first order dynamics and homogeneous damping. 
We now consider a realistic model of the European high-voltage transmission network [\textit{PanTaGruEl}~\cite{pagnier2019pantagruel}] with second order dynamics on generator nodes (i.e., $m_i>0$), and first order dynamics on load nodes (i.e., $m_i=0$). 
Moreover, generator and load parameters are defined to mimic the real European grid and thus inertia and damping parameters are inhomogeneous over the network. 
In such systems, time scales of line disturbances range from slow perturbation in malfunctioning transformers, to fast ones due to re-closing attempts, e.g., following lightning strikes or groundings.

Figure~\ref{fig:pantagruel}(a) shows the success rate of $\bm{\psi}$-based and $\bm{x}$-based methods. 
As expected, at low disturbance frequencies, the $\bm{\psi}$-based method performs well, while the $\bm{x}$-based method is not able to distinguish disturbed nodes. 
Interestingly, when the disturbance frequency $\Omega$ increases and reaches the system spectrum, the success rate of the $\bm{\psi}$-based method is reduced, but it still successfully locates the disturbed line in a majority of cases. 
This performance loss is presumably due to the fact that, when the disturbance frequency $\Omega$ enters the range of the system's eigenfrequencies, system's modes can be excited, leading to large oscillations in chunks of the system and consequently rendering the location more challenging. 
Nonetheless, if one is able to catch the disturbance in its very first few cycles, i.e., before it has the chance to significantly excite any eigenmode, then the $\bm{\psi}$-based method is correct in more than 84\% of cases. 
We observe that for this particular application only the $\bm{\psi}$-based method is of use, the $\bm{x}$-based one becomes efficient only for frequencies that are way above what is of physical significance for power systems (i.e., $\Omega\gg 500$Hz). 
Conveniently, Fig.~\ref{fig:pantagruel}(b) shows that, for both approaches, the confidence closely follows the success rate. 
When success rate decreases, the confidence values get lower and more spread. 

Finally, in Figure~\ref{fig:pantagruel}(c), we are interested in knowing how far from a successful location the method is when it fails. 
To measure this, we introduce the following rank  
\begin{align}
r &= \max(k,l)\, ,\label{eq:rank}
\end{align}
where $(i_k,i_l)$ is the disturbed line, using the indexing defined at Eq.~\eqref{eq:ordering}. 
Obviously, $r=2$ in case of a successful location. The average rank is only slightly over $2$, with an average rank of $2.3$ and $3.4$ for early (after one perturbation cycle) and late (after a few tens of perturbation cycles) detection respectively. Furthermore, the early detection has a worse-case rank of $7$. Hence, even when the method fails, it can still be a useful tool to guide the human operator by reducing the set of plausible causes of the disturbance to only a few elements.

\section{Conclusion}
The main lesson from the above observations is that, even though the frequency mismatch is less accurate in identifying a fast perturbed line (as can be expected from an analytical perspective), it proves to perform almost as good as the direct agents trajectories. 
For first order dynamics, the $\bm{\psi}$-based approach can hardly be improved in our examples. 
Therefore, if one had to choose a single indicator to detect and locate a line disturbance, the frequency mismatch would be a good candidate. 
Naturally, computing the frequency mismatch $\bm{\psi}(t)$ requires to know the agents positions $\bm{x}(t)$, and then knowledge of the latter come for free in our method. 
Hence, by combining the time series of $\bm{x}$ and $\bm{\psi}$, one is guaranteed to be able to locate the disturbed line, regardless of the disturbance time scale. 

For second order dynamics, the systems eigenmodes appear to have a much more dramatic effect on the inference. 
By relying on early measurement (i.e., shortly after the disturbance kicks in), the inference performs reasonably well. 
Furthermore, in our example, the $\bm{\psi}$-based approach is the only efficient one over the practically relevant time scales. 

Finally, we conjecture that, in the case of multiple, simultaneous perturbations, the frequency mismatch will correctly identify the disturbed lines, provided they are not too close to each other. 
Indeed, our previous work [\cite{delabays2021locating}] showed that the frequency mismatch clearly identifies multiple line disturbances. 
Furthermore, as noticed above, fast perturbations do not propagate throughout the network and therefore do not interfere in the frequency mismatch.

\end{document}